\documentclass[%
 aip,
 amsmath,amssymb,
 reprint,%
]{revtex4-1}

\usepackage{graphicx}
\usepackage{dcolumn}
\usepackage{bm}
\usepackage{physics}
\usepackage[utf8]{inputenc}
\usepackage[T1]{fontenc}
\usepackage{mathptmx}

\begin{document}


\title[Polariton Frequency Generator]{High-Frequency Exciton-Polariton Clock Generator}

\author{C. Leblanc}
\affiliation{Institut Pascal, University Clermont Auvergne, CNRS, SIGMA Clermont, F-63000 Clermont-Ferrand, France} 
\author{G. Malpuech}
\affiliation{Institut Pascal, University Clermont Auvergne, CNRS, SIGMA Clermont, F-63000 Clermont-Ferrand, France} 
\author{D. D. Solnyshkov}
\affiliation{Institut Pascal, University Clermont Auvergne, CNRS, SIGMA Clermont, F-63000 Clermont-Ferrand, France}
\affiliation{Institut Universitaire de France (IUF), F-75231 Paris, France}
 \email{dmitry.solnyshkov@uca.fr}

\date{\today}

\begin{abstract}
Integrated circuits of photonic components are the goal of applied polaritonics. Here, we propose a compact clock generator based on an exciton-polariton micropillar, providing optical signal with modulation frequency up to 100 GHz. This generator can be used for driving polariton devices. The clock frequency can be controlled by the driving laser frequency. The device also features low power consumption (1 pJ/pulse).
\end{abstract}

\maketitle

Optical computing is an important long-standing goal in the field of photonics \cite{Ambs2010,Touch2017}. Different approaches were used in the field during the last 70 years, including analog optical computing \cite{Fabre2014,Solli2015}, combined electro-optical circuits \cite{Ying2018}, photonic neural networks \cite{Woods2012}, and digital optical computing \cite{Sawchuk1984,Touch2017b,Wherrett1995}. Integration of the photonic components raises several key problems linked with miniaturisation, such as the protection from the parasite reflections, potentially solved by the recently emerged topological photonics \cite{Lu2014,Ozawa2019}. But one of the main problems has been linked with the high powers required for optical switching via the Kerr nonlinearity or other mechanisms used in nonlinear optics \cite{Rozanov1997,Reinisch1994,Zheludev1989}. One option to solve this issue can be to reduce the size of the non-linear cavities, such as it is done for nano-lasers \cite{hamel2015spontaneous}. Another option is to enhance non-linearities using the strong light-matter coupling \cite{Hopfield1958, Microcavities} which allows to use the significant interactions between the matter part of the eigenmodes \cite{Ciuti98,Vladimirova2010,sun2017direct}.

Cavity exciton-polaritons \cite{Microcavities} (polaritons) are a good example of such promising platform showing an increase by a factor of $10^4$ with respect to the ordinary Kerr nonlinearity in standard inorganic semiconductors \cite{Paraiso2010,sun2017direct} as well as in 2D monolayer materials \cite{barachati2018interacting,emmanuele2019highly} and perovskites  \cite{fieramosca2019two}. Thanks to these properties, many non-linear polariton devices have already been proposed and implemented \cite{gao2012polariton,ballarini2013all,Nguyen2013,Sanvitto2016,zasedatelev2019room}. Such devices exhibit low operation powers and fast switching times. They can be assembled into logical circuits capable of functioning at very high operating frequencies. However, in order to be correctly tested and to be ultimately useful, these circuits have to be driven not by an external pulsed (or even \emph{cw}) laser, as it is  typically the case in experiments, but by an integrated "clock generator" able to provide the expected operation frequency. 

In electronics, the clock generators always contain a non-linear element (such as an inverter), which is often combined with a resonant element (such as a quartz crystal) for frequency stability. The most well-known example is the Pierce oscillator \cite{Pierce1923}. In general, the non-linear circuits used for generating oscillations date back to the beginning of the XX century \cite{Eccles1919} and are called multivibrators. They are often based on bistable nonlinear elements exhibiting two possible stationary outputs for a given single input. Biasing the bistable element makes impossible for it to remain in these stationary states, and thus the element is constantly switching between the two at a well defined rate. It is therefore logical to apply this well-developped approach in photonics. Bistable elements showing low switching powers have been extensively studied in polaritonic systems \cite{Baas2004,Paraiso2010,Gippius2007}, but not in the oscillatory regime. Several theoretical works considering oscillations involving polariton bistability in 2 coupled pillar cavities have appeared in the recent years \cite{Sarchi2008,Solnyshkov2009}, but none of them focused on the precise purpose of high-frequency clock signal generation.

In this work, we lay the cornerstone for the future polariton computer. We design and optimize a high-frequency polariton clock generator that could be used in various configurations to drive the optical logical devices. The generator has the simplest possible configuration: a single polariton micropillar with an output channel. Coherent nonlinear oscillations based on the parametric scattering  have already been observed in a single polariton pillar \cite{Ferrier2010}, which makes the implementation of our proposal quite realistic. Before complete polariton circuits are built, the clock generator can be used for testing the switching performance of various nonlinear elements. The operating frequency of such elements is found of the order of a few hundreds GHz, close to the THz gap, which makes difficult the use of other types of generators, for example, electro-optical ones.

For relatively low wave vectors, the polariton wave function  $\psi(r,t)$ is well described by a nonlinear Schr\"odinger equation (Gross-Pitaevskii equation) with pump and decay. To simplify the initial theoretical description, we neglect the non-parabolicity of the polariton dispersion and the polarization degree of freedom:
\begin{equation}
    i\hbar\frac{\partial \psi}{\partial t}=-\frac{\hbar^2}{2m}\Delta \psi+\alpha \left|\psi\right|^2\psi +U\psi-\frac{i\hbar}{2\tau}\psi+Pe^{-i\omega t}
    \label{GPE}
\end{equation}
where $m=5\times 10^{-5}m_0$ is the typical polariton mass ($m_0$ is the free electron mass), $\alpha=5~\mu$eV$\mu$m$^2$ is the coefficient of polariton nonlinear interactions \cite{Ferrier2011,Nguyen2015}, $\tau$ is the polariton lifetime (varying from few ps to few hundred ps), $U(\mathbf{r})$ is the confinement potential obtained by lithography (few tens of meV scale), and $P(\mathbf{r})$ is the pump. Its frequency $\omega$ is measured with respect to the energy of a free polariton (approximately 1500 meV in this material system). This is a full 2D model, which we shall use for the numerical simulations of the system. The chosen parameters are typical for GaAs samples.

The system we consider, described by the potential $U(x,y)$, consists of a polariton micropillar of a circular cross-section (radius $R$). 
A single mode in presence of a quasi-resonant pump and non-linearity is known to be bistable \cite{Baas2004}, but it cannot exhibit a complex behavior, such as oscillations. To obtain such behavior, at least two modes are required. This can be obtained either by considering two degenerate modes of coupled pillars \cite{Sarchi2008,Abbarchi2013} or two non-degenerate modes of a single pillar \cite{Ferrier2010}. It is the latter configuration that we choose because of its simplicity. The cylindrical pillar is supposed to be pumped exactly in the center with a relatively small pumping spot, optimized for the overlap with the two pumped states.

\begin{figure}[tbp]
    \centering
    \includegraphics[width=0.8\linewidth]{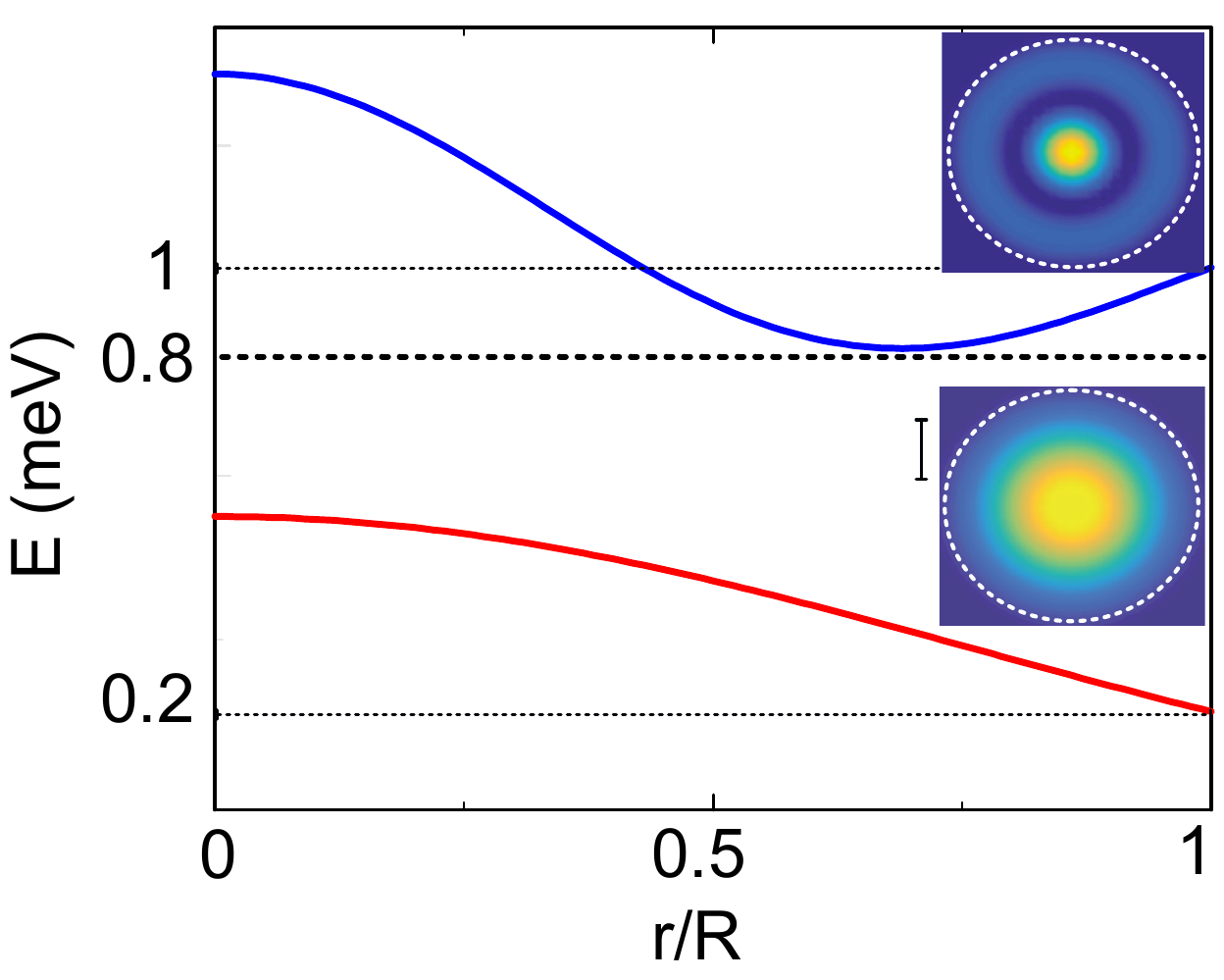}
    \caption{Energies (dotted lines) and wave functions (solid lines) of the two lowest-energy symmetric modes of a micropillar. A possible position of the laser energy is shown as a dashed line. Inset: spatial images of the two states, with the pillar boundary shown as a white dashed circle.}
    \label{fig1}
\end{figure}

In order to obtain a qualitative understanding of the behavior of the system, we decompose the global condensate wavefunction $\psi(x,y,t)$ on the basis of the quantized modes of the micropillar $\psi_n(x,y)$. These modes are solutions of the stationary eigenvalue equation $\hat{H}_0\psi_n=E_n\psi_n$, where $\hat{H}_0$ is the conservative part of Eq.~\ref{GPE} with the potential $U$ restricted to a single pillar (without the output channel). The decomposition writes: 
$\psi(x,y,t)=\sum_n c_n(t)\psi_n(x,y)e^{-iE_nt/\hbar}$, 
where the complex coefficients $c_n(t)$ define the populations of the corresponding modes and their emission intensity. Projecting the nonlinear Schr\"odinger equation~\eqref{GPE} on the eigenstates gives a system of coupled equations for $c_n(t)$. The coupling appears because of the interacting term $\bra{\psi_m}\alpha|\psi|^2\ket{\psi_n}\neq 0$, mixing the eigenstates. We need to take into account only the modes that are supposed to be strongly populated, that is, the modes exhibiting a large overlap integral with the pump. The modes of a cylindrical pillar can be written as $\psi_{n,l}=\chi_{n,|l|}J_{|l|}(kr)e^{il\phi}$, where $n,l$ are the radial and azimuthal quantum numbers, $J$ is the Bessel function of the first kind, $k=\hbar^{-1}\sqrt{2mE_{n,l}}$, $\chi$ is a normalization constant, and the energies are determined by the zeros of the Bessel function: $E_{n,l}=\hbar^2 j_{n+1,l}^2/2mR^2$ ($j_{n+1,l}$ is the $n+1$st zero of the Bessel function $J_{l}$). Since the pumping is symmetric, all wave functions with nonzero $l$ have a vanishing overlap with it. The two states of interest, the closest to the bottom and thus the easiest to obtain and to observe, shall be denoted as $\psi_{a}=\psi_{0,0}$ and $\psi_{b}=\psi_{1,0}$ (see Fig.~\ref{fig1}). The projection of the nonlinear term $|\psi|^2$ can be simplified by analysing the expected behavior. Indeed, only the ground state can exhibit a bistable jump and become strongly populated, because $E_a<\hbar\omega$. In such conditions, the contribution of the other mode to the coupling term can be neglected. The coupling term therefore reads
\begin{equation}
    \bra{\psi_a}\alpha|\psi|^2\ket{\psi_b}=\alpha |c_a|^2 \iint \psi_a^* |\psi_a|^2 \psi_b~dr d\phi=J|c_a|^2
\end{equation}
The sign of $J$ depends on the relative phase of the modes, which, in turn, depends on their actual position with respect to the laser. The two modes are going to have opposite phases because $E_a<\hbar\omega<E_b$ (approximately $\pi$ phase difference if the decay is neglected), and thus $J<0$.  This can be understood qualitatively as follows: the admixture of the upper state brings the ground state closer to the Thomas-Fermi limit \cite{Pitaevskii}. Numerical evaluation of the Bessel functions gives $JS\approx 1.3\alpha$.
For the nonlinear shift in the energy of each mode, we neglect the contribution of the other mode and of the interference terms, because of the smaller overlap integrals. 

\begin{figure}[tbp]
    \centering
    \includegraphics[width=1\linewidth]{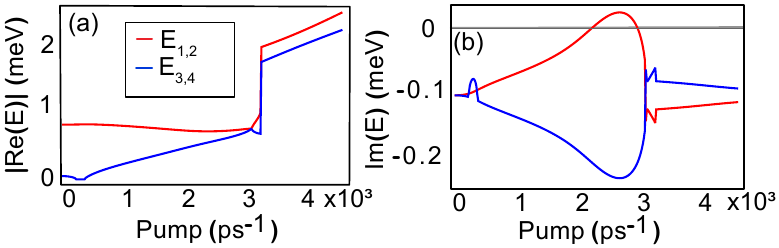}
    \caption{Real (a) and imaginary (b) parts of the energy of weak perturbations of the stationary solution for a pillar of $R=4.5$ $\mu$m, a polariton lifetime $\tau=7$ ps and a laser frequency of $\hbar\omega=0.9$ meV.}
    \label{fig2}
\end{figure}

With all these approximations, the coupled mode equations can be written as
\begin{eqnarray}
\label{0dt}
i\hbar\frac{\partial c_a}{\partial t}&=& \left(\hbar\omega_a +\alpha_a\left|c_a\right|^2-\frac{i\hbar}{2\tau}\right)c_a+J\left|c_a\right|^2 c_b+P_ae^{-i\omega t}\\
i\hbar\frac{\partial c_b}{\partial t}&=& \left(\hbar\omega_b +\alpha_b\left|c_b\right|^2-\frac{i\hbar}{2\tau}\right)c_b+J\left|c_a\right|^2 c_a+P_be^{-i\omega t}\nonumber
\end{eqnarray}
To simplify the analysis, we then assume $\alpha_a=\alpha_b=\alpha_0$ (the self-interactions in the 2 modes are approximately the same) and $P_a=P_b=P_0$ (the overlap of the pump with the 2 modes is almost the same). In order to identify the necessary conditions for the oscillations, we then perform a stability analysis. We begin by writing the system~\eqref{0dt} in the stationary limit, when the only frequency in the system is the laser frequency: $\psi_{a,b}(t)=\psi_{a,b}^s e^{-i\omega t}$:
\begin{equation}
\label{0dstat}
 \left(\hbar(\omega_{a,b}-\omega) +\alpha_0\left|c_{a,b}^s\right|^2-\frac{i\hbar}{2\tau}\right)c_{a,b}^s+J\left|c_a^s\right|^2 c_{b,a}^s+P_0=0\nonumber
\end{equation}
Following the Bogoliubov-de Gennes approach, we study the weak perturbations of the stationary solution $\psi_{a,b}(t)=e^{-i\omega t}\left(\psi_{a,b}^s+U_{a,b}e^{-iEt/\hbar}+V_{a,b}^*e^{iE^*t/\hbar}\right)$ which results in a secular equation allowing to find the eigenenergies $E$ of these excitations. Positive imaginary part $\Im E>0$ of any eigenvalue means that the corresponding perturbation grows exponentially, and the real part of its energy is a good estimation for the frequency of the self-sustained oscillations $\omega_{osc}\approx\Re E/\hbar$,  once a new regime settles down. Figure~\ref{fig2} shows the real and imaginary parts of the energies of weak perturbations as a function of pumping $P_0$ (parameters given in the caption). We see that there is a range of pumping values (between $2\times10^3$ and $3\times10^3$~particles/ps) which correspond to the growth of the perturbations $\Im E>0$. It is in this range of parameters that our device is expected to operate. Moreover, since the real part of the energy is almost constant in this range, the generator is well-protected from the fluctuations of the driving laser intensity.

The approximate frequency of the generated clock oscillations can be found numerically from the solution of the 4th order secular equation, but in order to have a clear analytical estimate, we need to reduce the system size, keeping only the important terms: the energies of the two modes (taking into account the effect of the bistability) and their coupling. For this estimate, we assume that the lower mode jumps to the level of the pump, which determines both its energy and its occupancy. The reduced 2-coupled modes Hamiltonian reads:
\begin{equation}
{\hat H_{red}} = \left( {\begin{array}{*{20}{c}}
{\hbar \omega}&{J\frac{{\hbar \omega  - {E_a}}}{\alpha }}\\
{J\frac{{\hbar \omega  - {E_a}}}{\alpha }}&{{E_b}}
\end{array}} \right)
\end{equation}
which allows to find the frequency of the oscillations as $\nu=\sqrt{(E_a-\hbar\omega)^2+4(E_b-\hbar\omega)^2J^2/\alpha^2}/2h$ (the factor $1/2$ appears because we analyze the oscillations of intensity). For a pillar of $R=4.5~\mu$m with the mode energies of $E_a=0.2$~meV and $E_b=1$~meV, and a laser frequency of $\hbar\omega=0.9$~meV, this expression predicts a frequency of $86$~GHz, very close to the value of $91$~GHz predicted by the numerical solution of the Bogoliubov-de Gennes equations.

\begin{figure}[tbp]
    \centering
    \includegraphics[width=1\linewidth]{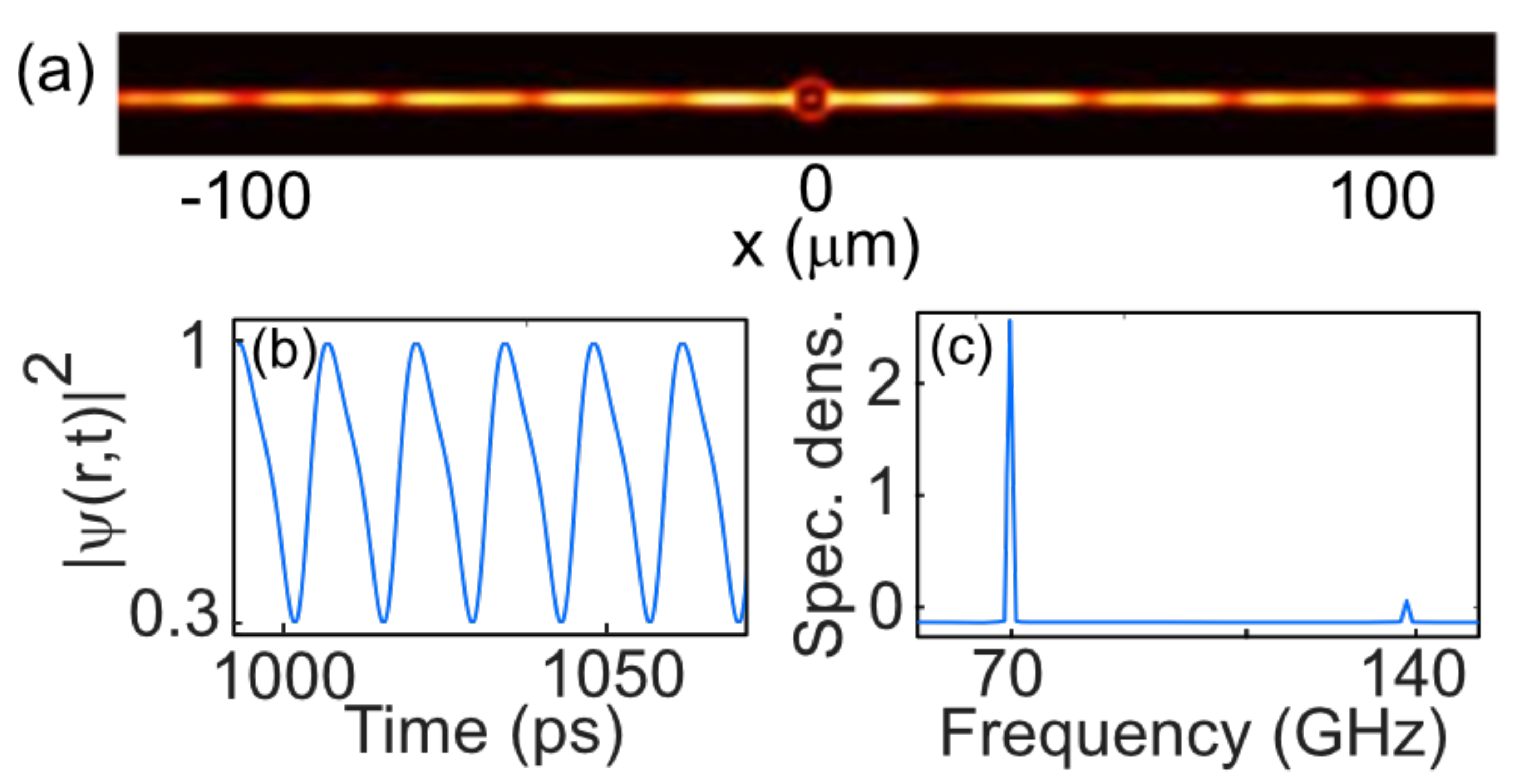}
    \caption{(a) Spatial polariton density snapshot from a 2D numerical simulation. Propagating density pulses can be observed in the channels. (b) Oscillations of the polariton density for a pillar of $R=5$ $\mu$m, a laser frequency of $\hbar\omega=0.82$ meV and a pumping of $p=2000$~ps$^{-1}$. (c) Spectral density of $|\psi(r,t)|^2$ showing a resulting frequency of $\nu=70$ GHz.}
    \label{fig3}
\end{figure}

In order to check our analytical predictions, we perform two sets of numerical simulations. The first set is based on a full 2D Gross-Pitaevskii equation \eqref{GPE} and the second is based on the two-mode approximation \eqref{0dt}. The results of one 2D simulation  is shown in Fig.~\ref{fig3} (parameters are given in the caption). Panel (a) shows the spatial distribution of intensity in the system composed of a circular pillar with 2 output channels. The visible periodic spatial modulation of intensity in the channels is due to the propagating periodically oscillating signal. Fig.~\ref{fig3}(b) shows the time dependence of the polariton density for a fixed point in a channel, far from the pillar and the boundary of the system. High-contrast stable periodic oscillations are clearly visible. These oscillations set in during the initial transitional period of 1-2 hundreds of ps, and remain stable for at least several dozen of nanoseconds. 

The corresponding spectral density is shown in panel \ref{fig3}(c). The main frequency is 70~GHz, very close to the simple analytical estimate of 74~GHz provided by the equation above. This high oscillation frequency is obtained for realistic parameters, as we discuss below.

\begin{figure}[tbp]
    \centering
    \includegraphics[width=1\linewidth]{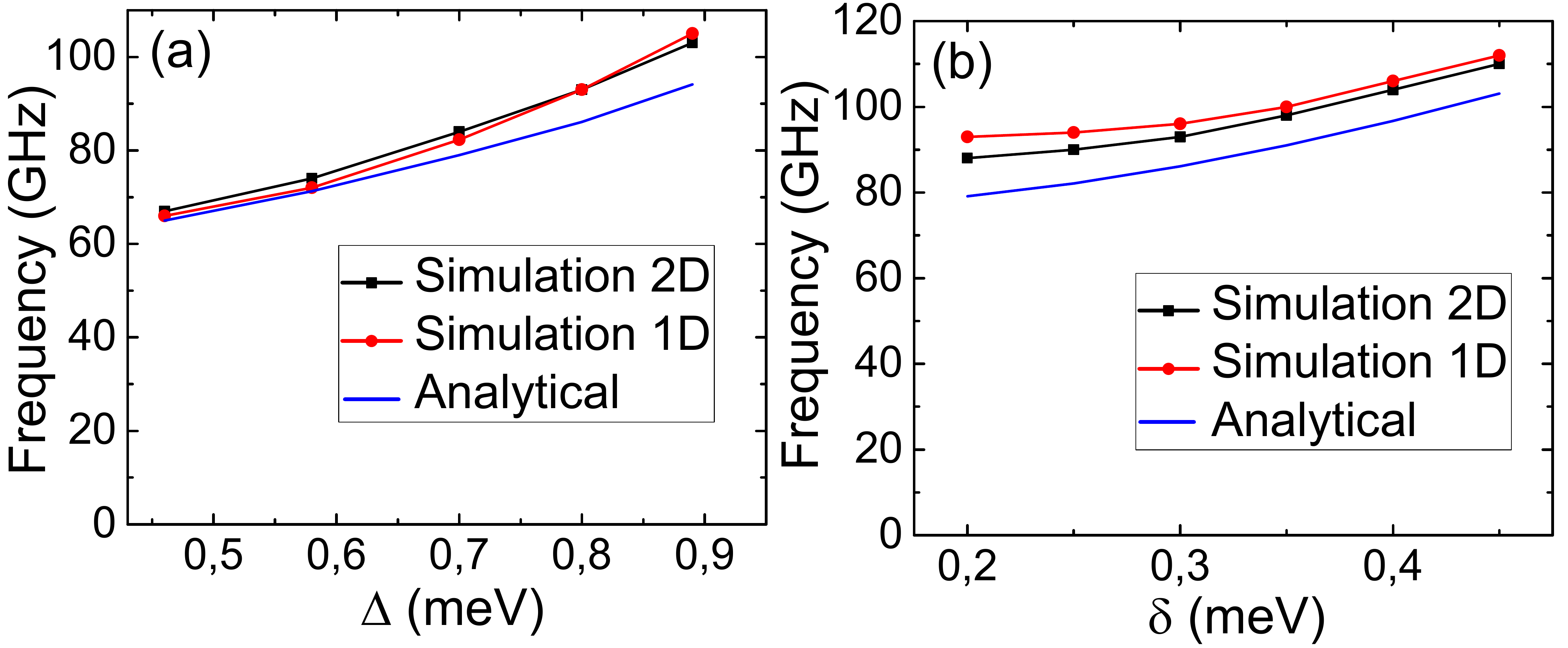}
    \caption{(a) Frequency as a function of $\Delta$ for $\delta=0.3$ meV constant. (b) Dependence of the frequency with $\delta$ for $\Delta=0.8$ meV constant. The pumping varies between 1500 and 3000 ps$^{-1}$.}
    \label{fig4}
\end{figure}

Figure~\ref{fig4} presents the dependence of the most practically important parameter -- the clock frequency -- on the parameters of the system, which can be varied experimentally. These parameters include the energy difference between the two modes of the pillar $\Delta=\hbar(\omega_3-\omega_1)$ (determined by the size of the pillar) and the laser detuning with respect to the middle frequency   $\delta=\hbar\omega-\hbar(\omega_1+\omega_3)/2$. As an example, for a pillar of $R=4.5$~$\mu$m and a laser frequency of $\hbar\omega=0.9$~meV, we obtain $\Delta=0.8$~meV and $\delta=0.3$~meV. Fig.~\ref{fig4}(a) shows the increase of the frequency from 70 to above 100~GHz with the increase of mode splitting $\Delta$  (for a constant detuning $\delta=0.3$~meV), meaning that the smaller is the pillar, the higher is the oscillation frequency. The detuning $\delta$ of the laser is also an important parameter in order to optimize the output frequency, as shown in Fig.~\ref{fig4}(b). Unfortunately, both $\Delta$ and $\delta$ cannot be increased indefinitely, because the efficiency of the laser pumping decreases with its detuning with respect to the mode. For a mode linewidth of $\gamma\approx0.1$~meV, the maximal experimentally realistic offset is $\sim 7\gamma\approx0.7$~meV \cite{Nguyen2015}. It means that the realistic values of $\delta$ in Fig.~\ref{fig4}(b) are below $0.3$ meV. Nevertheless, we provide the results for higher values of $\delta$, which might be accessible for broader pillar resonances. These two figures confirm that the operation frequency can be tuned in a broad range of values. The most accessible way for tuning the clock generator is via the pumping laser frequency.

We stress that in the 2D simulations the value of the lifetime is different in the pillar (7~ps$^{-1}$) and in the channels (300~ps$^{-1}$), which can be easily achieved experimentally
. This is important, because it is the decay rate $\gamma$ which constrains the maximal operation frequency via $\delta$ and $\Delta$. Sufficiently high decay rate is required for high-frequency operation. We also note that the generator frequency is relatively stable with respect to the pumping power: it changes only by 10\% within the whole range of pumping powers where the oscillations can be observed (for constant $\Delta$ and $\delta$). Besides the frequency, another important parameter of the clock generator is the power consumption. The power of the optical pump used to drive the system is of the order of 10~$\mu$W. This is approximately 10 times higher than in the best modern electronic clock generators. However, since the operation frequency of the present device is much higher, energy consumption per pulse is of the order of   1~pJ, much better than in electronics.

To conclude, we have studied the possibility of fabrication of a clock generator for integrated polaritonic circuits based on a single polariton micropillar. Such device demonstrates high generation frequencies of about 100~GHz, required to fully exploit the potential of polariton circuits for ultra-fast data treatment. We have shown that the operation frequency can be controlled by the laser detuning (in-situ) or by changing the pillar radius.

\begin{acknowledgments}
 We thank S. Porteboeuf-Houssais for useful comments. We acknowledge the support of the project "Quantum Fluids of Light"  (ANR-16-CE30-0021), and of the ANR program "Investissements d'Avenir" through the IDEX-ISITE initiative 16-IDEX-0001 (CAP 20-25). 
\end{acknowledgments}

\bibliography{references}

\end{document}